\documentclass[final,5p,times,twocolumn]{elsarticle}

\usepackage{amsmath,amscd,amssymb}
\usepackage{bm}
\usepackage{graphicx}
\usepackage{units}
\usepackage{setspace}
\usepackage{lineno}
\usepackage{color}
\usepackage[colorlinks=true,linkcolor=blue, citecolor=blue]{hyperref}%
\usepackage{siunitx}

\renewcommand{\vec}[1]{\bm{#1}}

\journal{Ultrasonics}

\begin{document}
\begin{frontmatter}

\title{Designing single-beam multitrapping acoustical tweezers}%
\author{Glauber T.\ Silva\thanks{Corresponding author: \texttt{glauber@pq.cnpq.br}} and Andr\'e L. Baggio}
\cortext[cor1]{Corresponding author: \texttt{glauber@pq.cnpq.br}}
\address{Physical Acoustics Group, Instituto de F\'isica, Universidade Federal de Alagoas, Macei\'o, AL 57072-970, Brazil.}

\begin{abstract}
 The  concept of a single-beam acoustical tweezer device which can simultaneously trap  microparticles at
 different points is proposed and demonstrated through computational simulations.
 The  device employs an ultrasound beam produced by a circular focused transducer operating at $\unit[1]{MHz}$ 
 in water medium.
 The ultrasound beam exerts a radiation force that may tweeze suspended microparticles in the medium. 
 Simulations show that the acoustical tweezer can simultaneously trap microparticles 
 in the  pre-focal zone along the beam axis, i.e. between the transducer surface and its geometric focus.
 As acoustical tweezers are fast becoming a  key instrument in microparticle handling,
 the development of acoustic multitrapping concept may turn into a useful tool in engineering these devices.
\end{abstract}


\end{frontmatter}

\section{Introduction}
Noncontact  particle handling methods based on optical~\cite{ashkin1970} and
acoustic~\cite{torr1984} radiation forces are promoting 
a revolution in biotechnology and biomedical applications~\cite{Grier2003,laurell2007,friend2011,ding2012,li2013}.
Techniques for particle trapping  which employ  a laser and an ultrasound beam are
known, respectively, as optical tweezer~\cite{ashkin1986} and acoustical tweezer~\cite{wu1991}.
Methods based on acoustical tweezers are potentially useful in applications for which optical tweezers
can hardly operate.
For instance, handling particles in opaque medium to electromagnetic radiation cannot be performed by
optical tweezers.
Furthermore, a laser beam may damage biological structures by heating
and also by a process called photodamage, which is related to the formation of singlet oxygen  when photon absorption occurs~\cite{neuman1999}.

Two different approaches have been employed to acoustically trap microparticles, namely
standing waves and single-beam methods.
The first acoustical tweezer device used two counterpropagating focused 
ultrasound beams to form a standing wave at $\unit[3.5]{MHz}$, which was used 
to trap  $\SI{270}{\micro\meter}$-diameter latex particles
and frog eggs in water~\cite{wu1991}.
In other arrangement, an ultrasound standing wave generated between a $\unit[2.1]{MHz}$ linear array and a reflector 
trapped alumina microparticles with $\SI{16}{\micro\meter}$ mean diameter~\cite{kozuka1998}.
Devices in the acoustofluidics realm are, in general, based on standing waves~\cite{evander2012,Glynne-Jones2012}.
Standing surface acoustic waves (SSAW) have been used to trap particles with diameter smaller than 
$\SI{8}{\micro\meter}$ 
suspended in microfluidic channels~\cite{shi2009,tran2012}.
Standing Bessel waves were generated by a circular $64$-element ultrasonic array  to manipulate
$\SI{90}{\micro\meter}$-diameter polystyrene microparticles in 2D~\cite{courtney2014}.
Tilted standing waves produced by a  three 
PZT transducer setup which operated at $\unit[1.67]{MHz}$ were used
to  trap and transport a $\SI{100}{\micro\meter}$-diameter silica bead~\cite{meng2014}.
Furthermore, the standing wave method has been also employed to levitate
particles and droplets in air~\cite{foresti2013,foresti2014}.

On the other hand, single-beam acoustical tweezers  utilize
tightly focused transducers and linear arrays for trapping microparticles in the device focus point.
In particular, a $\unit[30]{MHz}$-transducer with an f-number of $0.75$ tweezed 
a $\SI{126}{\micro\meter}$-diameter lipid microdroplet~\cite{lee2009}.
Higher frequency transducers operating at $\unit[200]{MHz}$ have been designed to handle a 
$\SI{10}{\micro\meter}$-diameter leukemia cell~\cite{lee2011} and microsphere~\cite{lam2013}.
A PZT transducer equipped with a multi-foci Fresnel lens generated a $\unit[17.9]{MHz}$ zeroth-order Bessel beam,
which was employed to trap microspheres ranging in diameter from $70$ to $\SI{90}{\micro\meter}$~\cite{choe2011}.
A $\unit[57.5]{MHz}$ needle hydrophone with an f-number of $1$ produced an ultrasound
be which tweezed a $\SI{30}{\micro\meter}$ diameter lipid microdroplet~\cite{hsu2012}.
Also, a $64$-element linear phased array operating at $\unit[26]{MHz}$ was able to
trap a $\SI{45}{\micro\meter}$-diameter polystyrene microparticle~\cite{zheng2012}.

The key aspect in designing  single-beam acoustical tweezers  is how to form a beam to 
trap a particle with an specific size.
Different schemes of acoustical tweezer beamforming have been previously 
investigated for a circular focused~\cite{wu1990, chen1996, baresch2013} and
a linear array~\cite{sapozhnikov2013} transducer.
In these studies, only one  trapping point located in the transducer focal zone was considered.
Recently, the possibility of trapping a particle in the nearfield of piezoelectric 
disk was discussed~\cite{mitri2013}.
Perhaps the most serious disadvantage of this method is that the
yielded ultrasound beam behaves like a plane progressive wave in the vicinity of the beam axis at the nearfield~\cite{pierce:book}.
Hence, the transverse radiation force associated to the beam may not be strong enough to hold a particle in 3D.
In this work, a  method to form multiple microparticle traps in the pre-focal region of a  piezoelectric 
focused transducer is proposed.
The method's ability to tweeze microparticles is theoretically demonstrated through computational simulations.
In so doing, the radiation force produced by a $\unit[1]{MHz}$ piezoelectric  transducer, with focus at
$\unit[50]{mm}$ and an f-number of $1.25$,
which is readily available commercially, is computed on a  microdroplet, 
made of either benzene or peanut oil, suspended in water. 
For this transducer configuration,  trapping points arise in the pre-focal zone.
After obtaining the radiation force field, the dynamics of the microparticle trapping is simulated considering
effects of gravity, buoyancy and Stokes' drag.
The results show that  microparticle entrapment occurs in points as close as one third of the transducer focal distance.

\section{Ultrasound beamforming}
\begin{figure}
\centering
 \includegraphics[scale=.65]{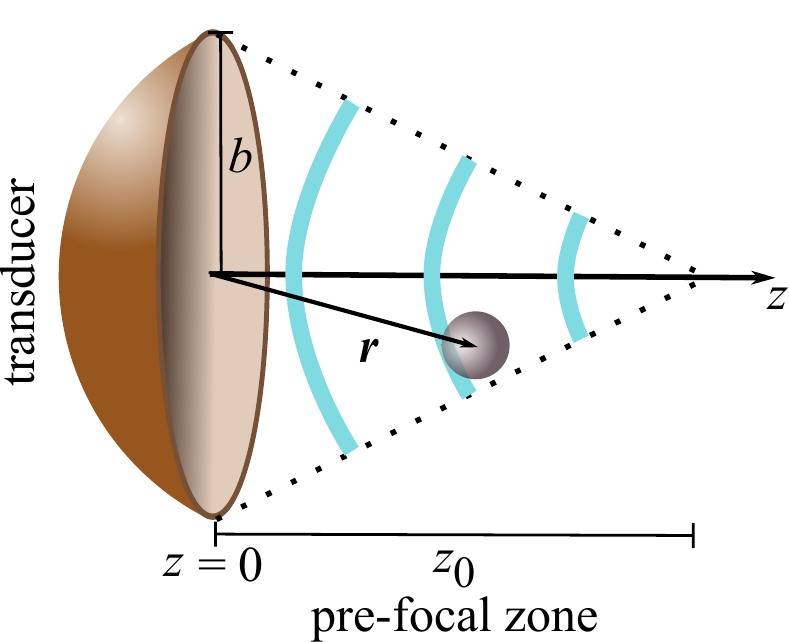}
 \caption{(Color online) Circular focused ultrasound transducer with aperture $2b$ and focus $z_0$ 
 actuating on a spherical particle of radius $a$ located at $\vec{r}$ in the medium.
 \label{fig:transducer}}
\end{figure}

Before calculating the acoustic radiation force exerted on a particle,
we need to establish the ultrasound beamforming by the focused transducer.
Hence, consider  a circular focused  transducer, with aperture $2b$ and curvature radius $z_0$,
 mounted on a compliant baffled at  $z=0$ (see Fig.~\ref{fig:transducer}).
The transducer is immersed in a inviscid fluid of density $\rho_0$ and speed of sound $c_0$ 
and is uniformly excited with a sinusoidal signal of angular frequency $\omega$.
Thus, the  ultrasound beam produced by the transducer is a time-harmonic wave described by 
its pressure $p_\textrm{in}(\vec{r}) \textrm{e}^{-\textrm{i} \omega t}$  and  velocity  of a fluid parcel 
$\vec{v}_\textrm{in}(\vec{r}) \textrm{e}^{-\textrm{i} \omega t}$,
both observed at the time $t$ in the position $\vec{r}$ in a fixed coordinate system.
Hereafter, the time-harmonic term $\textrm{e}^{-\textrm{i} \omega t}$ will suppressed for the sake simplicity.
A useful pressure-velocity relation in first-order approximation is obtained from the momentum conservation
equation as follows~\cite{landau1993}
\begin{equation}
\vec{v}_\textrm{in}= -\frac{\text{i}}{\rho_0 c_0 k}\nabla p_\textrm{in},
\label{vp}
\end{equation}
where $k=\omega/c_0$.

It is further assumed that the transducer concavity is fairly small and
the wavelength is much smaller than its radius.
Hence the ultrasound beam can be described in the paraxial approximation.
In terms of the transducer parameter, these assumptions read~\cite{lucas1982}
\begin{equation}
\label{paraxial}
 \frac{1}{8N^2}\ll1 \text{ and } 1\ll kb \ll 128 \pi N^3,
\end{equation}
where $N=z_0/(2b)$ is the transducer f-number.
The model discussed here may not be suitable for transducers with $N<1$.
On that account a different beamforming model might be considered~\cite{mitri2014}.

By assuming that the transducer generates an axisymmetric paraxial beam,
the pressure field can be expressed in cylindrical coordinates $(\varrho, z)$ as
\begin{equation}
\label{p_rhoz}
 p_\textrm{in}(\varrho,z) = -\text{i}\rho_0 c_0 k e^{\text{i}k z} q(\varrho,z),
\end{equation}
where $q$ is the velocity potential characteristic function.
For a circular focused transducer mounted on a perfectly compliant baffle at $z=0$,
one can show that the characteristic velocity potential is given by~\cite{lucas1982}
\begin{align}
\nonumber
 q(\varrho, z) &= 
\frac{v_0}{z} \exp\left( \frac{\text{i} k \varrho^2 }{2z}\right)
\int_0^b  \exp\left[ \frac{\text{i} k \varrho'^2 }{2} \left(\frac{1}{z} - \frac{1}{z_0} \right)\right]\\
& \times J_0\left( \frac{k \varrho \varrho'}{z}\right) \varrho' \text{d}\varrho'.
 \label{qrz2}
\end{align}
where $v_0$ is the magnitude of the normal vibration  velocity on the transducer surface,
 and $J_m$ is the $m$th-order Bessel function.
In the focal plane $z=z_0$, Eq.~(\ref{qrz2}) becomes 
\begin{equation}
 q(\varrho,z_0) = -b v_0 \frac{J_1(k b \varrho/z_0)}{k \varrho}.
 \label{qfocal}
\end{equation}

%The function $q$ satisfies the parabolic wave equation
%\begin{equation}
%\label{eq_parabolic}
% \nabla^2_\perp q+ 2 \text{i} k \partial_z q = 0,
%\end{equation}
%where $\nabla_\perp$ is the transverse gradient and $\partial_z=\partial/\partial z$
In the paraxial approximation,
it follows from Eq.~(\ref{vp}) and (\ref{p_rhoz}) that the fluid  velocity is
\begin{equation}
\vec{v}_\textrm{in} \approx -\left[
(\partial_\varrho q)\vec{e}_\varrho 
+ \text{i} k q \vec{e}_z \right]  e^{\text{i}k z},
\label{vp2}
\end{equation}
where $\vec{e}_\varrho$ and  $\vec{e}_z$ are the unit-vector along the radial direction and  $z$-axis.
Here we are using the shorthand notation $\partial_\varrho =\partial/\partial \varrho$ and
$\partial_z =\partial/\partial z$.
Note that in deriving Eq.~(\ref{vp2}),  we have neglected the term $\partial_z q$,
because $|\partial_z q| \ll k |q|$.
This assumption is valid in a region not very near to the transducer surface, say $z>0.3 z_0$.

\section{Radiation force on a Rayleigh particle}
Our attention now turns to the acoustic radiation force exerted by the  ultrasound beam on a particle
much smaller than the wavelength, i.e. the so-called  Rayleigh regime.
The particle 
has radius $a$, density $\rho_1$,   speed of sound $c_1$, and
its position is  denoted by $\vec{r}$. 
Viscous effects of the host fluid in the radiation force analysis are neglected.
This hypothesis holds when the external viscous boundary layer of the particle~\cite{landau1993} 
$\delta_0 = \sqrt{2 \nu_0/\omega}$ is
much smaller than the particle radius,
where $\nu_0$ is the kinematic viscosity of the host fluid.

Using the radiation force formulas in Cartesian coordinates given in Ref.~\cite{silva2011}, one can show that
the acoustic radiation force $\bm{F}^\textrm{rad}$ on the Rayleigh particle is the sum of three 
components~\cite{silva2014}, 
namely the gradient $\vec{F}^\text{grad}$, the scattering 
$\vec{F}^\text{sca}$, and the absorption $\vec{F}^\text{abs}$ radiation forces.
Thus, we have
\begin{equation}
\bm{F}^\textrm{rad} = \vec{F}^\text{grad} + \vec{F}^\text{sca} + \vec{F}^\text{abs}.
\end{equation}

The gradient radiation force is given by~\cite{gorkov1962}
\begin{equation}
\label{grad}
 \vec{F}^\text{grad}( \vec{r}) = - \nabla U^\textrm{rad}( \vec{r}),
\end{equation}
where
\begin{equation}
\label{potential_function}
 U^\textrm{rad} =   \pi a^3  \left(f_0 \frac{|p_\textrm{in}|^2}{3 \rho_0 c_0^2} - f_1 \frac{\rho_0 |\vec{v}_\textrm{in}|^2}{2}\right)
\end{equation}
is the radiation force potential energy.
The quantities $f_0 = 1 - \rho_0 c_0^2/(\rho_1 c_1^2)$ and
$f_1 = 2(\rho_1- \rho_0)/(2 \rho_1 + \rho_0)$
are the compressibility and density contrast factors of the particle, respectively.
The gradient radiation force appears due to the interference between the incident
and the scattered waves.
Moreover, this force is responsible for trapping particles at
the minima of  the potential energy $U^\textrm{rad}$.

The scattering radiation force is given by~\cite{silva2014}
\begin{align}
\nonumber
 \vec{F}^\text{sca}(\vec{r}) &=  \pi a^2 (ka)^4\biggl[ \frac{4  }{9 }\left(f_0^2 + f_0f_1 + \frac{3f_1^2}{4} \right) 
  \frac{\overline{\vec{I}}(\vec{r})}{c_0}\\
  &-  \frac{f_1^2 }{6 k}\textrm{Im}[\nabla \cdot \rho_0{\bm v}_\textrm{in} {\bm v }_\textrm{in}^*({\bm r})]
 \biggr],
 \label{fscat1}
\end{align}
where $\overline{\vec{I}}= (1/2)\textrm{Re}[p_\textrm{in} \vec{v}_\textrm{in}^*]$ is the 
 incident  intensity averaged in time and $\rho_0{\bm v}_\textrm{in} {\bm v }_\textrm{in}^*$ is
 a dyadic (second-rank tensor).
 
 The absorption radiation force reads~\cite{silva2014}
\begin{align}
\nonumber
 {\bm F}^\textrm{abs}(\vec{r}) &=  \pi a^2 \tilde{\alpha}_1 k a\biggl[\frac{8(1-f_0) }{3}    
    \frac{ \overline{\bm I}(\vec{r})}{c_0} - \frac{12(1-f_0)}{5( \tilde{\rho}^{-1}_1 + 2)^2} (ka)^2  \\
  &\times \frac{1}{k} \textrm{Im}[\nabla \cdot \rho_0{\bm v}_\textrm{in} {\bm v }_\textrm{in}^*(\vec{r})]
    \biggr],
    \label{fabs}
\end{align}
where $\tilde{\rho}_1=\rho_1/\rho_0$ and 
$\tilde{\alpha}_1= \alpha_1 (\omega/2 \pi)^2/k_1$ is the normalized absorption of the particle,
with $\alpha_1$ being the absorption 
coefficient and $k_1=\omega/c_1$.
This equation is valid when absorption inside the particle is described through
the  dispersion relation
\begin{equation}
\kappa_1 = k_1 (1 + \textrm{i} \tilde{\alpha}_1).
\end{equation}
Furthermore, we also assume that the inner viscous boundary layer~\cite{epstein1953} 
$\delta_1 = \sqrt{2 \nu_1/\omega}$ is much smaller than the particle radius, where $\nu_1$ 
is the kinematic viscosity of the particle.

The absorption and the scattering radiation forces are only relevant when the beam has a traveling part,
i.e. it locally behaves as a traveling plane wave.
In the focal region, the acoustic beam has a traveling part.
Thus, in principle, all three components of the radiation force in the focal region have to be
taken into account.
Both absorption and scattering radiation force point to the forward beam propagation 
direction.
Obviously, these components tend to break the axial trapping stability.

The  contribution of the momentum flux  divergence
$\textrm{Im}[\nabla \cdot \rho_0{\bm v}_\textrm{in} {\bm v }_\textrm{in}^*] $ 
and the time-averaged intensity $ \overline{\bm I}$ depend on the incident  beam.
In Appendix A, expressions of these two quantities are derived in the axial direction 
for a focused ultrasound beam in the paraxial approximation.
In this case, we antecipate that for the transducer parameters which will be consider later,
the contribution of the momentum flux divergence to the absorption and scattering
radiation forces is much smaller than that due to the averaged intensity.

In the focal plane, a particle might be trapped in $\varrho=0$ if $\partial_\varrho^2 U^\textrm{rad}(0,z_0) >0$.
Using Eqs.~(\ref{p_rhoz}), (\ref{qfocal}) and (\ref{vp2}) into Eq.~(\ref{potential_function}),
one can show that this condition is satisfied when
\begin{equation}
f_0< - \frac{3(k^2 b^4 + 64 z_0^2)}{8 (k b z_0)^2}f_1\approx  - \frac{3f_1}{32N^2}.
\label{trap}
\end{equation}
If the particle is denser than the host fluid ($\rho_1>\rho_0$), then $0\le f_1\le 1$.
Thus, to have a stable trap in the transducer focus it is necessary that $f_0<0$
which means that the particle should be less compressible than the host fluid.
Yet not proven here, this behavior has been observed in the radiation force computed
in the transducer pre-focal zone.

After obtaining the radiation force,
one cannot assure that the possible trapping points are stable.
In fact, entrapment also depends on  gravity, buoyancy, and Stokes' drag acting on the microdroplet.
Assume that gravity and buoyancy lies along the $y$-axis.
Thus, the effective energy potential on the particle is given by
\begin{equation}
\label{Ueff}
U^\text{eff} = U^\textrm{rad} + \frac{4 \pi a^3}{3} (\rho_1-\rho_0) g y,
\end{equation}
where $g=\unit[9.8]{m/s^2}$ is the Earth gravity acceleration.

\section{Microdroplet dynamics}
The microdroplet dynamics is computed considering the dynamic viscosity of water  $\mu_0=\unit[10^{-3}]{Pa\cdot s}$.
Let $\vec{r}$ and $\dot{\vec{r}}= d\vec{r}/dt$ denote, respectively, 
the  position and the velocity of a microdroplet at the time $t$.
The flow around the microdroplet is assumed to be laminar.
Hence, the drag  force on the microdroplet is given by the Stokes' law~\cite{landau1993} 
$\vec{F}^\text{drag} = 6 \pi \mu_0 a \dot{\vec{r}}$.
According to Newton's second law, the microdroplet position  is described by
the  differential equation
\begin{equation}
\ddot{\vec{r} } + \gamma \dot{\vec{r}} = \vec{a}^\text{rad}(\vec{r}) + \vec{g}^\text{eff},
\label{particle_dyn}
\end{equation}
where $\gamma = 9  \mu_0 / (2 a^2 \rho_1 )$ is viscous damping coefficient,
$\vec{a}^\text{rad} = 3 \vec{F}^\text{rad} / (4 \pi a^3 \rho_1)$
is the  acceleration due to the radiation force, and
$\vec{g}^\text{eff} = (\rho_0/\rho_1 -1) g \vec{e}_y$ is 
the acceleration due to gravity and buoyancy effects,  with $\vec{e}_y$ 
being the unit-vector along the $y$-axis.
Equation~(\ref{particle_dyn}) was solved using the forth-order Runge-Kutta method in
\textsc{Matlab} (The MathWorks Inc., Natick MA, USA).

The particle dynamics discussed here neglects acoustic streaming
caused by nonlinear and absorption effects in the host medium.
In fact, acoustic streaming may affect entrapment owing to the fact
that it may cause an additional drag force on the trapped particle.
However, a study found that acoustic streaming by a focused source is more significant
in the focal and pos-focal regions~\cite{kamakura1995}.

\section{Numerical results and discussion}
Consider the host medium as water with the following physical parameter
$\rho_0=\unit[1000]{kg/m^3}$, $c_0=\unit[1480]{m/s}$, and $\nu_0=\unit[10^{-6}]{m^2/s}$.
The  circular focused  transducer used to generated the incident ultrasound beam to the microdroplets
 operates at $\unit[1]{MHz}$ with intensity $I_0= \rho_0 c_0 v_0^2/2=\unit[3.3]{kW/m^2}$.
At this frequency, the external viscous boundary around the microdroplet is $\delta_0 = \SI{0.56}{\micro \meter}$.
The transducer has aperture $2b=\unit[40]{mm}$, focal distance $z_0=\unit[50]{mm}$, and an f-number $N=1.25$.
Since $kb=84.9$ and $N=1.25$, the paraxial approximation inequalities in (\ref{paraxial}) are readily satisfied.  
Two types of microdroplets of radii $a=\SI{58.9}{\micro \meter}$ and size parameter $ka=0.25$ (Rayleigh particles) 
made of benzene and peanut oil are considered for trapping.
Note that the microdroplets are immiscible in water.
At room temperature, the physical parameters for benzene  are~\cite{kino1987}
$\rho_1=\unit[870]{kg/m^3}$, $c_1=\unit[1295]{m/s}$, $f_0=-0.5$, $f_1=-0.09$,
$\nu_1=\unit[7.47\times 10^{-7}]{m^2/s}$,
$\delta_1=\SI{0.46}{\micro \meter}$,
and
$\alpha_1 = \unit[8.73\times 10^{-13}]{Np\: MHz^{-2}\: m^{-1} }$;
while for  peanut oil we have~\cite{coupland1997} $\rho_1=\unit[913]{kg/m^3}$, $c_1=\unit[1465.9]{m/s}$, $f_0=-0.12$,
$f_1=-0.06$, $\nu_1=\unit[8.1\times 10^{-5}]{m^2/s}$,
 $\delta_1=\SI{5}{\micro \meter}$,
and $\alpha_1 = \unit[1.5\times 10^{-12}]{Np\: MHz^{-2}\: m^{-1}  }$.
With these parameters, the external and the internal viscous boundary layers of the microdroplets  satisfy 
$\delta_0\ll a$ and $\delta_1\le a/6$, respectively.

\subsection{Paraxial approximation model}
\begin{figure}[t]
\centering
 \includegraphics[scale=.5]{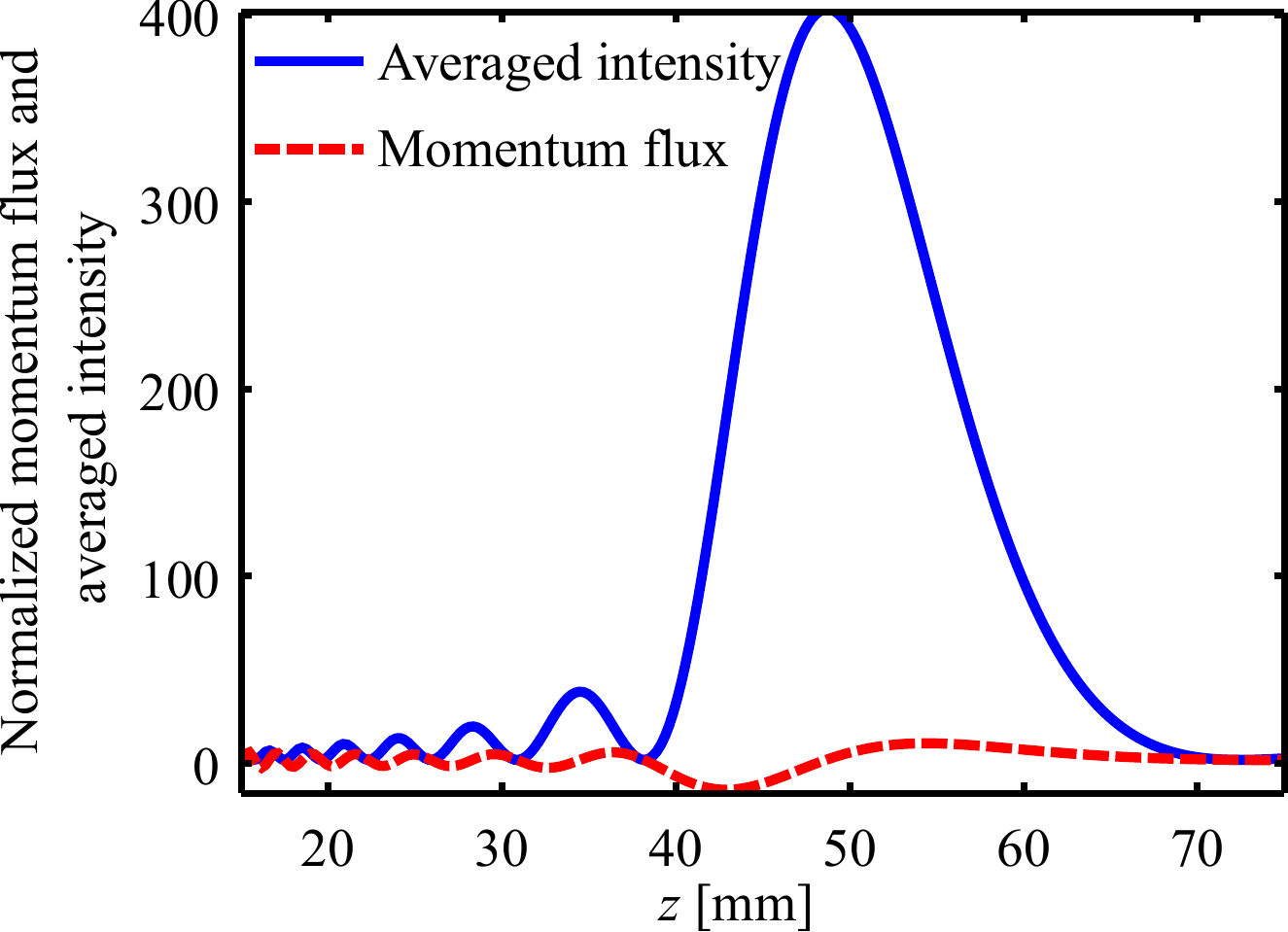}
 \caption{(Color online) 
 Normalized time-averaged intensity $\overline{\bm{I}}/I_0$ and   normalized momentum flux divergence 
 $k^{-1}c_0\textrm{Im}[\nabla \cdot \rho_0{\bm v}_\textrm{in} {\bm v }_\textrm{in}^*]/I_0$ along the $z$-axis,
 produced by  the focused transducer operating at $\unit[1]{MHz}$ frequency.
 \label{fig:flux_intensity}}
\end{figure}

In Fig.~\ref{fig:flux_intensity}, we show the normalized momentum flux divergence 
$k^{-1} c_0\textrm{Im}[\nabla \cdot \rho_0{\bm v}_\textrm{in} {\bm v }_\textrm{in}^*]/I_0$ and
the normalized
 time-averaged intensity $\overline{\bm{I}}/I_0$   along the $z$-axis.
These quantities were computed using Eqs.~(\ref{div_flux}) and (\ref{Iz2}).
Within the range $0.3 z_0<z<1.5 z_0$, the amplitude of the normalized momentum flux is less than $10\%$ of
that due to the normalized time-averaged intensity. 
Therefore, we may neglect the contributions of the momentum flux divergence term in 
both scattering and absorption radiation forces given in Eqs.~(\ref{fscat1}) and (\ref{fabs}), respectively.
\begin{figure}[t]
\centering
 \includegraphics[scale=.5]{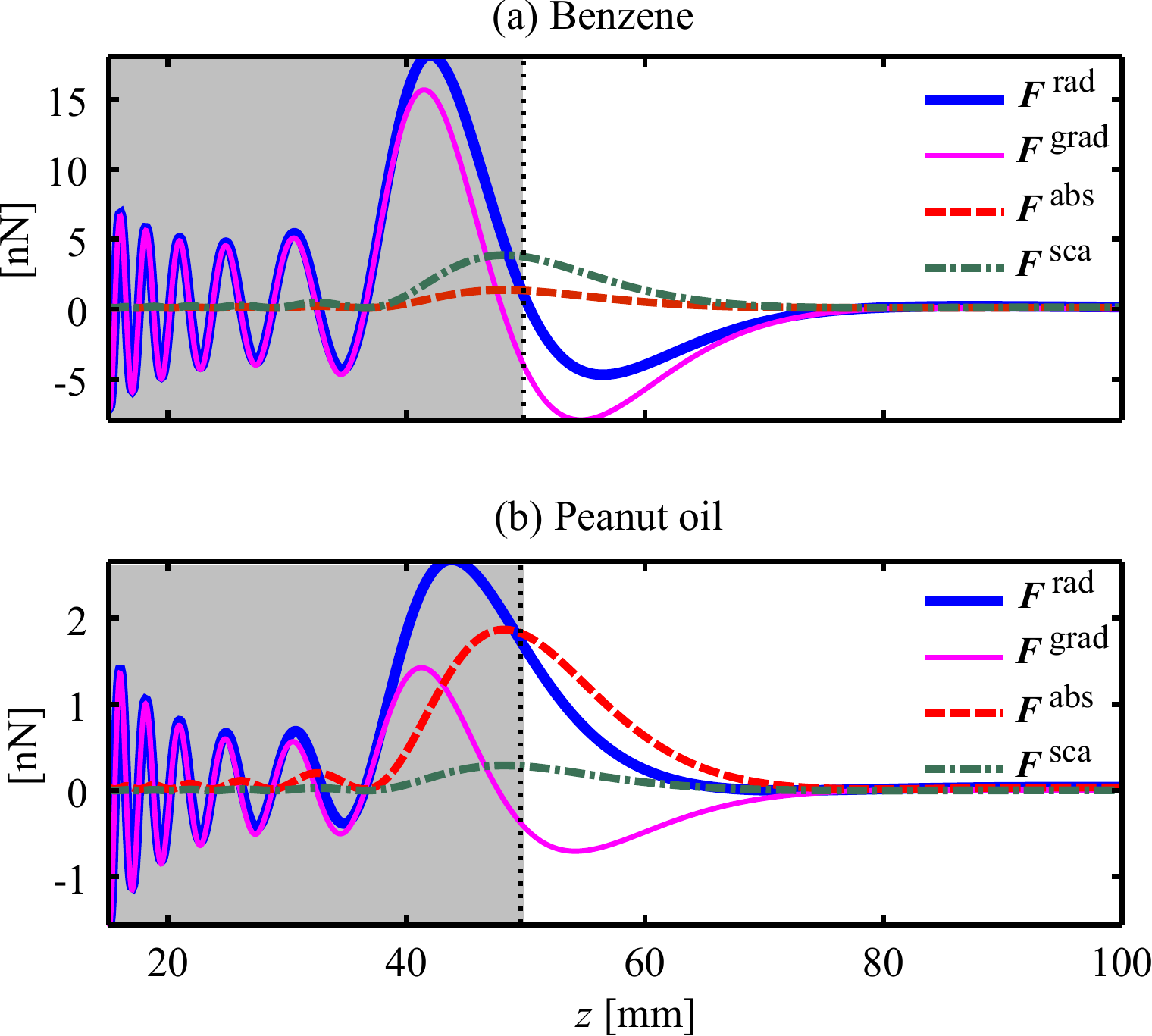}
 \caption{(Color online) Axial radiation force components produced by the focused transducer on  (a) a benzene
 and (b) a peanut oil microdroplet with size parameter  $ka=0.25$ and suspended in water.
 The transducer operates at $\unit[1]{MHz}$ with intensity $I_0=\unit[3.3]{kW/m^2}$.
 The radiation components were computed using Eqs.~(\ref{grad})-(\ref{fabs}).
 The region in gray depicts the transducer pre-focal zone $z<z_0=\unit[50]{mm}$.
 \label{fig:plot_axial}
 }
\end{figure}

The axial radiation force components acting on a benzene and a peanut oil microdroplet are shown in Fig.~\ref{fig:plot_axial}. 
The radiation components were computed using Eqs.~(\ref{grad})-(\ref{fabs}).
The gray region corresponds to the pre-focal zone of the transducer.
A trapping point is determined by observing that the radiation force has negative slope and
changes sign as a particle passes through it.
It is clear that the benzene microdroplet can be axially trapped at the  focal distance $z_0=\unit[50]{mm}$.
Nevertheless, this is not the case for the peanut oil microdroplet,  because
the absorption radiation force is greater than of the magnitude of the minimum value of the gradient  force.

\begin{figure}
\centering
 \includegraphics[scale=.5]{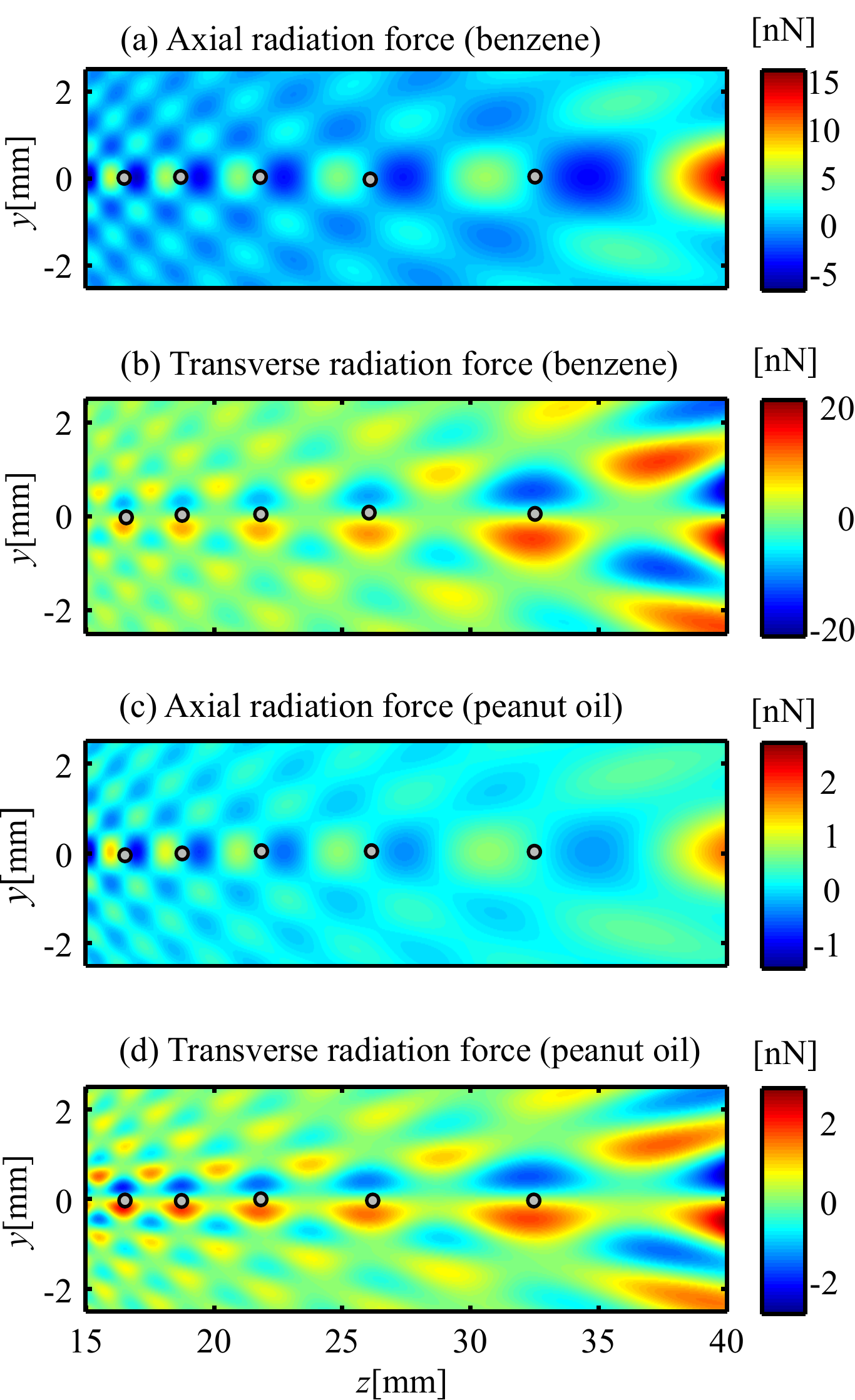}
 \caption{(Color online) The acoustic radiation force (axial and transverse) 
 produced by the focused transducer on (a) and (b) the benzene,
 and (c) and (d) the oil microdroplets with size parameter  $ka=0.25$ in water.
 The transducer operates at $\unit[1]{MHz}$ with intensity $I_0=\unit[3.3]{kW/m^2}$.
 Gray dots  indicate axial trapping points at $z=\unit[16.5, 18.8, 21.8, 26.1, 32.6]{mm}$.
 The acoustic radiation force was computed using  Eqs.~(\ref{grad})-(\ref{fabs}).
 \label{fig:axial_force}}
\end{figure}

In Fig.~\ref{fig:axial_force},  the axial and  transverse  radiation forces exerted on the microdroplets
are illustrated in the transducer pre-focal zone $z<z_0=\unit[50]{mm}$.
Five trapping points, marked with gray dots, are recognized in the pre-focal zone at 
$z=\unit[16.5, 18.8, 21.8, 26.1, 32.6]{mm}$.
It should be noticed that off-axial stable points are also present in the transducer pre-focal region.
\begin{figure}
\centering
 \includegraphics[scale=.5]{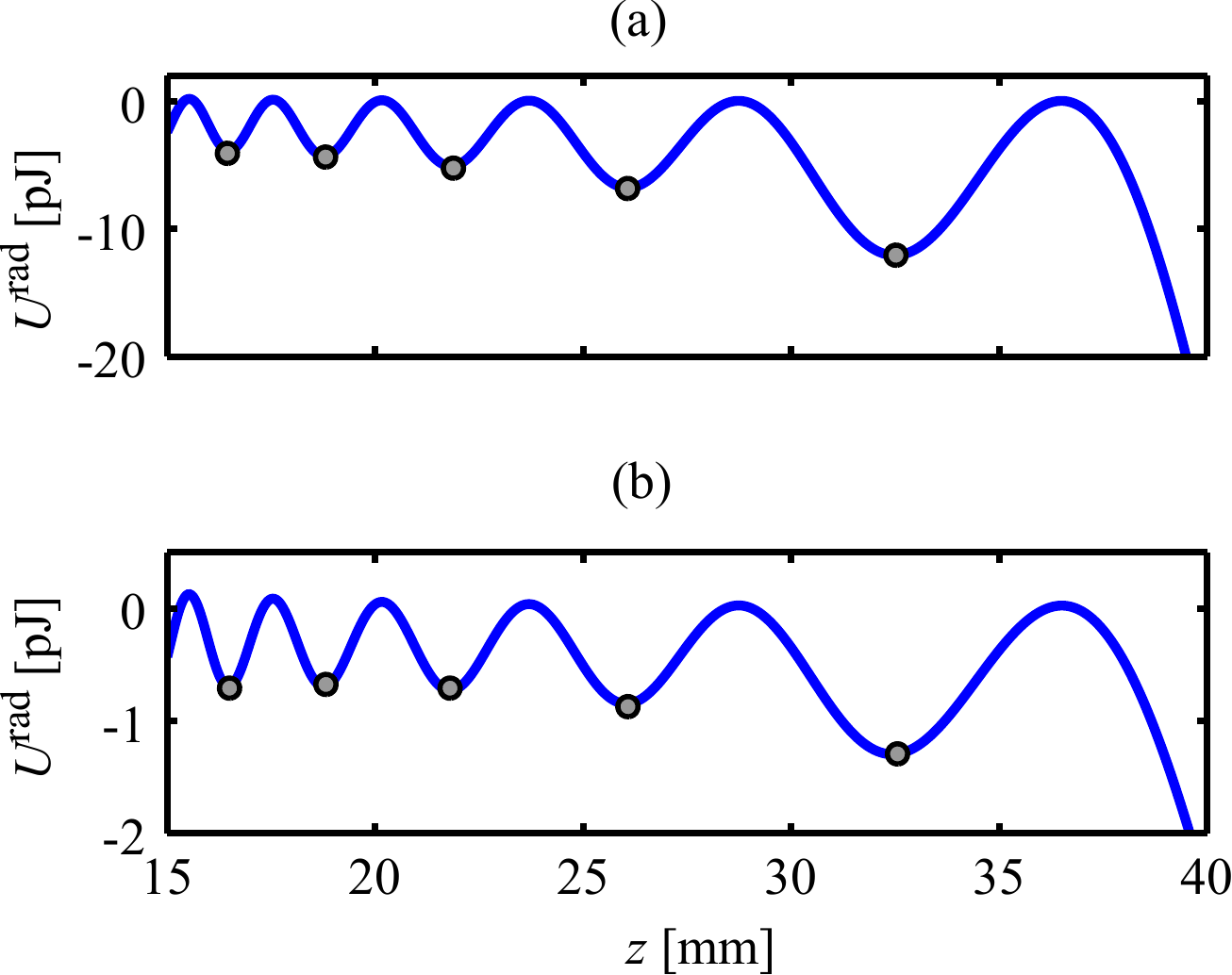}
 \caption{(Color online) Radiation force energy potential in the pre-focal zone $z<z_0=\unit[50]{mm}$
 generated by the focused transducer 
 for  (a) a benzene and (b) a peanut oil microdroplet with size parameter $ka = 0.25$ and suspended in water.
 The transducer operates at $\unit[1]{MHz}$ with intensity $I_0=\unit[3.3]{kW/m^2}$.
 The potenial was computed using Eq.~(\ref{potential_function}).
 The gray dots indicate the stable trapping points at $z=\unit[16.5, 18.8, 21.8, 26.1, 32.6]{mm}$.
 \label{fig:axial_potential}
 }
\end{figure}
\begin{figure}[t]
\centering
 \includegraphics[scale=.5]{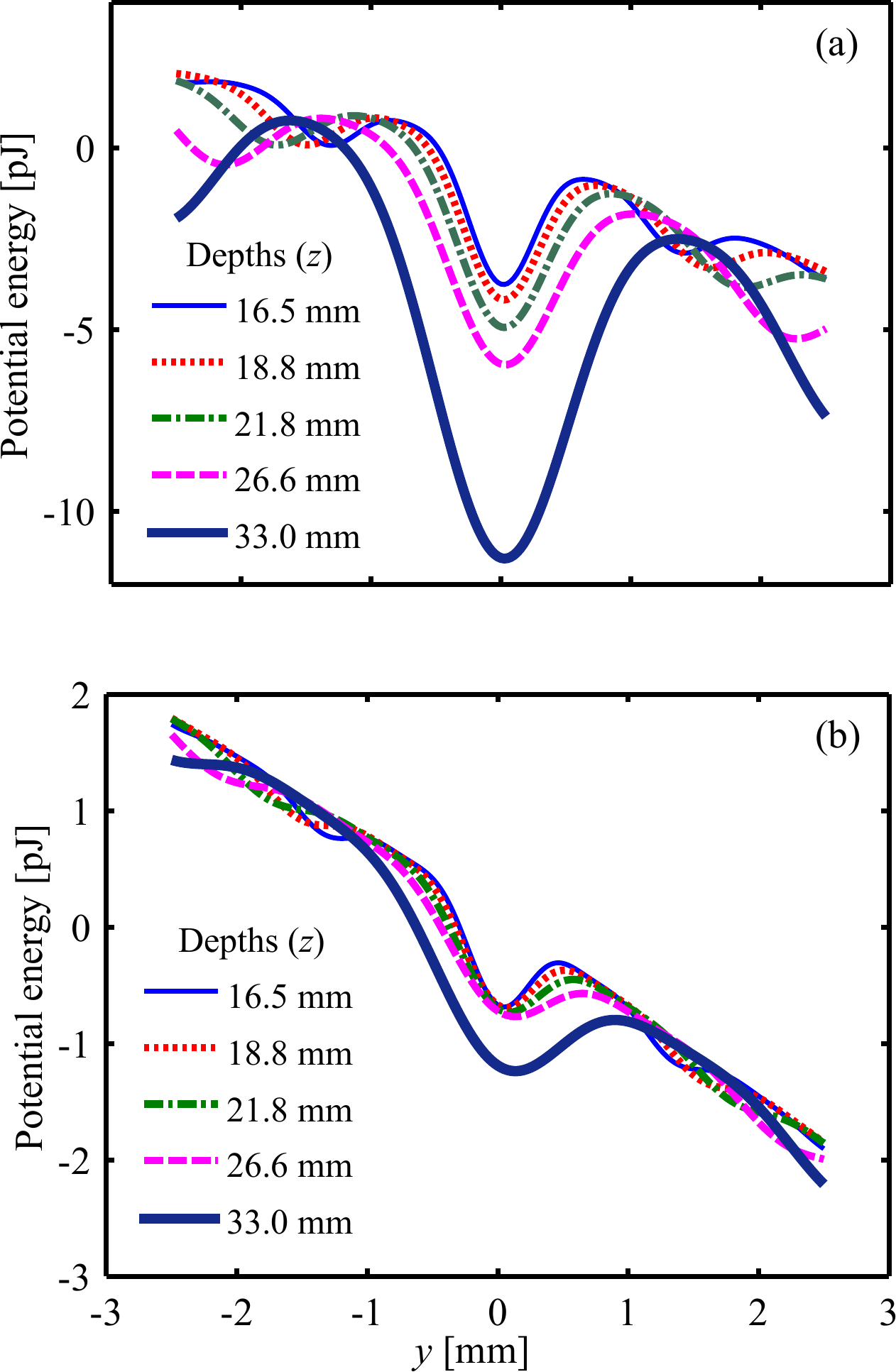}
 \caption{(Color online) Effective potential energy $U^\textrm{eff}$ along the $y$-axis of  
 (a) a benzene and (b) an  oil microdroplets with size parameter $ka = 0.25$ and suspended in water.
 The transducer operates at $\unit[1]{MHz}$ with intensity $I_0=\unit[3.3]{kW/m^2}$.
  The potenial functions were computed using Eq.~(\ref{potential_function}) at different depths
  corresponding to the gray dots in Fig.~\ref{fig:axial_potential}.
  \label{fig:pot_benzene_peanut}}
\end{figure}

In Fig.~\ref{fig:axial_potential}, we show the radiation force potential $U^\textrm{rad}$ along the $z$-axis in 
the transducer pre-focal zone $z<z_0=\unit[50]{mm}$.
The potential was calculated using Eq.~(\ref{potential_function}).
Five stable trapping points are noticed in the minima of the potential function.

The effective potential energy $U^\textrm{eff}$ along the $y$-axis of both microdroplets are depicted in Fig.~\ref{fig:pot_benzene_peanut}.
Clearly, potential wells are formed for all trapping points observed in Fig.~\ref{fig:axial_force}.
Thus, they are points of stable equilibrium.
The negative slope observed in the effective energy potential as $y$ increases 
appears because we considered gravity effects as described in Eq.~(\ref{Ueff}). 
If the microdroplet mechanical energy lies  within the potential well, then 
the microdroplet will oscillate around the minimum of
$U^\textrm{eff}$.
Due to the host fluid viscosity the mechanical energy is worn out
and the microdroplet is eventually trapped.
For the benzene microdroplet, the width of  the potential well  is about $50 a$ 
(where $a$ is the microdroplet radius)  at $z=\unit[32.6]{mm}$,
but it becomes $25 a$  at $z=\unit[16.5]{mm}$.
The width of the potential well for the oil droplet is $26a$ at $z=\unit[33]{mm}$,
whereas it  becomes $16 a$ at $z=\unit[16.5]{mm}$.
The obtained width of the potential wells suggests that many microparticles can be held on the same 
trapping point.
Moreover,  Figs.~\ref{fig:axial_potential} and \ref{fig:pot_benzene_peanut} show that the considered 
microdroplets can be three-dimensionally trapped in the medium.

\subsection{Microdroplet dynamics}

\begin{figure*}
\centering
 \includegraphics[scale=.5]{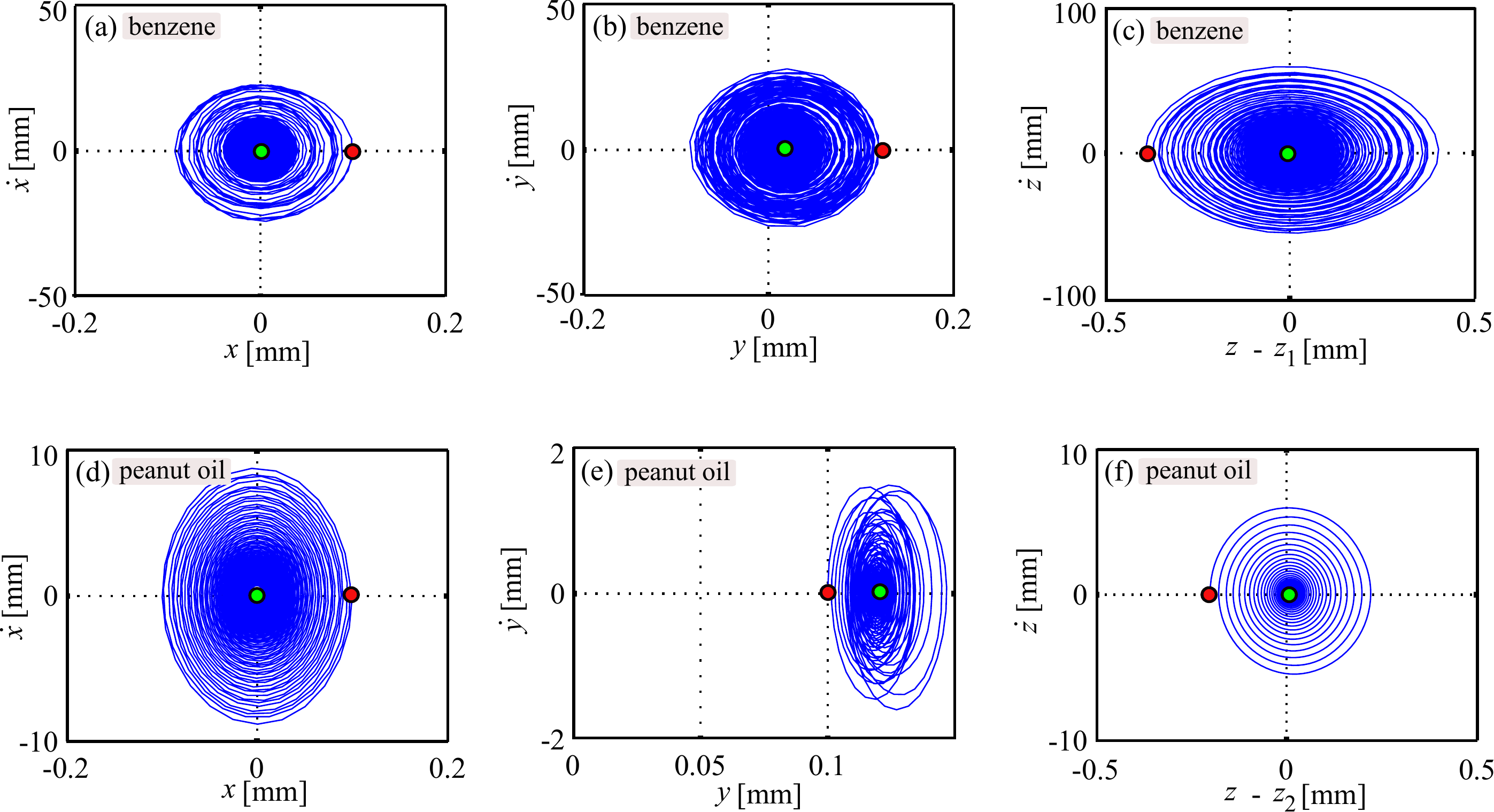}
 \caption{(Color online) The trajectories in phase space of (a)-(c)
    the benzene  and  (d)-(f) peanut oil  microdroplets suspended in water and 
    under the effective potential $U^\textrm{eff}$ given in Eq.~(\ref{Ueff}).
   The  trapping point  depths  are, respectively, at $z_1=\unit[16.5]{mm}$
   and $z_2=\unit[26.2]{mm}$.
   The red and the green dots indicate the initial and  final coordinates of the microdroplet, respectively.
  \label{fig:benzene_peanut_ps}}
\end{figure*}

The trajectories in phase space  of the  microdroplets are displayed in Fig.~\ref{fig:benzene_peanut_ps}.
The initial position of the benzene  and  oil microdroplets are, respectively,  $\unit[(0.1, 0.1, 16.1)]{mm}$ 
and $\unit[(0.1, 0.1, 26.0)]{mm}$, with zero initial velocity.
According to Fig.~\ref{fig:axial_force}, the nearest trapping points to the microdroplets initial positions 
are $z_1=\unit[16.5]{mm}$ and  $z_2=\unit[26.2]{mm}$.
It is worthy to note that the microdroplets dynamics is similar to a damped harmonic oscillator,
with a  characteristic decaying time (i.e. $5\%$ of the maximum displacement) of about $\unit[6]{s}$.
Beyond the decaying time, the microdroplets are  three-dimensionally trapped.
Referring to Fig.~\ref{fig:pot_benzene_peanut}, it is noticeable 
that the chosen trapping points have the shallowest potential wells in the pre-focal zone.
Even so, entrapment occurs.
Therefore, one should expect that the microdroplet will be held in all other trapping points,
since their associated potential wells are deeper than those chosen here.
Moreover, the particle dynamics in the $x$ and the $y$ directions are different
because gravity and buoyancy breaks the spatial symmetry of the radiation force potential $U^\textrm{rad}$.

\subsection{Finite element method}
Here we compared  the radiation force results obtained  in the paraxial approximation
as given in Eq.~(\ref{qrz2}), with computational simulations 
based on the finite element method (FEM).
The acoustic fields yielded by the circular focused
transducer driven at $\unit[1]{MHz}$ with intensity $\unit[3.3]{kW/m^2}$ was simulated
in COMSOL Multiphysics (Comsol Inc., Burlington, MA USA).
The perfect matched layer (PML)  boundary condition
was employed on the edges of the simulation domain, 
 to avoid spurious wave reflections back into the medium.

\begin{figure}
\centering
 \includegraphics[scale=.5]{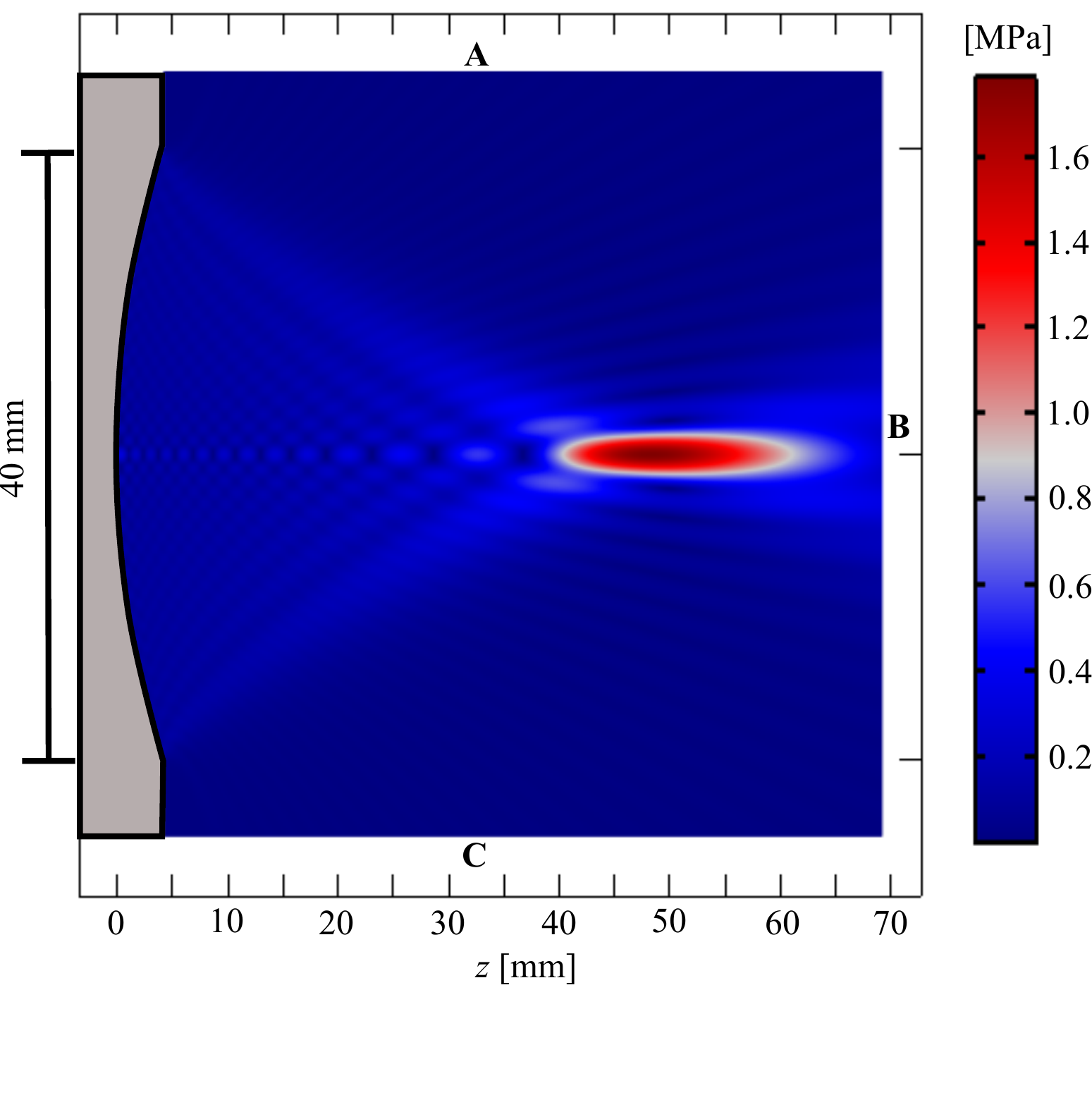}
 \caption{(Color online) Acoustic pressure produced by
 the circular focused transducer with aperture $2b=\unit[40]{mm}$,
 focus at $z_0=\unit[50]{mm}$,  $\unit[1]{MHz}$ frequency, and
 intensity of $I_0=\unit[3.3]{kW/m^2}$, operating in water at room temperature.
 The pressure in  obtained through computational 
 simulations based on the finite element method with
 perfect matched layer boundary conditions on the domain edges labeled as
 A,  B, and  C.
 The transducer is mounted on a rigid-baffle in which the fluid velocity is assumed
 to be zero.
 \label{fig:fem_pressure}}
\end{figure}

In Fig.~\ref{fig:fem_pressure}, we show the pressure generated by the transducer in
the FEM simulation.
The region describing water is formed by a triangular mesh having $8$ triangles per wavelength.
The simulation convergence was verified
by computing the spatially averaged pressure in a rectangular region of $4\times \unit[20]{mm}$ and
centered in the transducer focus point.
No significant change in the averaged pressure was observed by setting the number of triangles
per wavelength  larger than $4$.

\begin{figure}[t]
\centering
 \includegraphics[scale=.5]{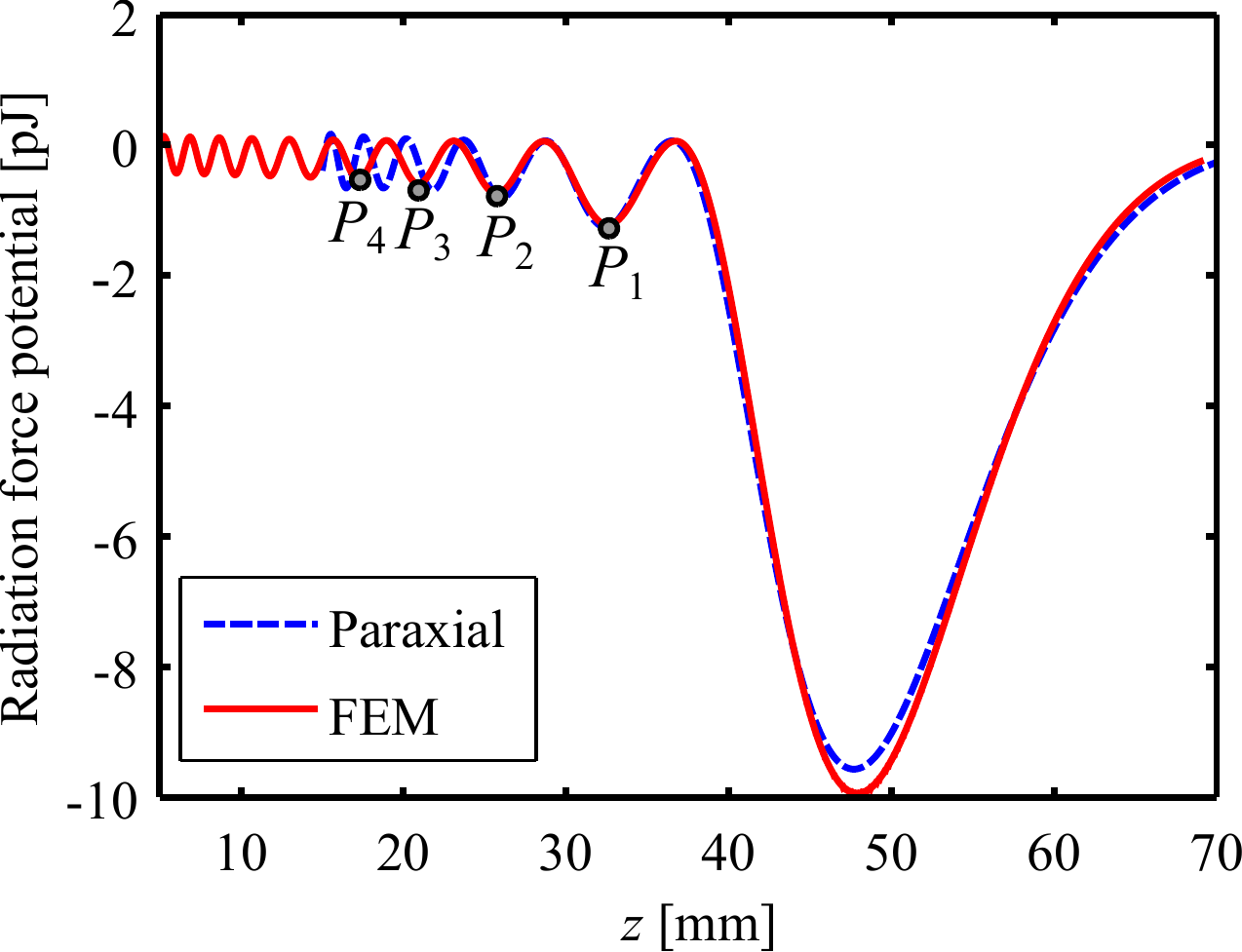}
 \caption{(Color online) The radiation force potential $U^\textrm{rad}$ of a peanut oil droplet of 
 size parameter $ka=0.25$, obtained with 
 the paraxial approximation and the finite-element methods.
 The points labeled as $P_i$ $(i=1,2,3,4)$ are located, respectively, at
 $z=32.6, 25.7, 20.9, \unit[17.2]{mm}$.
 \label{fig:fem_paraxial_Fgrad} }
\end{figure}

Because the acoustic radiation force in the transducer pre-focal zone is mostly
due to the gradient force, we computed the radiation force potential $U^\textrm{rad}$ of
an oil microdroplet using
Eq.~(\ref{potential_function}) for both paraxial approximation and finite element methods.
In Fig.~\ref{fig:fem_paraxial_Fgrad}, we show the radiation force potential
along the $z$-axis.
In the region between $15$ and $\unit[40]{mm}$, the FEM
simulation revealed four stable trapping points labeled as  $P_1,P_2,P_3,$ and $P_4$
at, respectively,  $z=32.6, 25.7, 20.9, \unit[17.2]{mm}$.
Points $P_1$ and $P_2$ coincide to those predicted by the paraxial approximation method.
Nonetheless the methods yield different locations for points $P_3$ and $P_4$.
The discrepancy between the two methods seen here
is somehow expected since the paraxial approximation  yields an unreliable solution in a region from $z=0$ to approximately 
$30\%$ of the transducer focal distance~\cite{lucas1982}.
In addition, the FEM simulations predict a total of nine axial trapping points in the pre-focal region.

\section{Summary and conclusions}
The concept of multitrapping acoustical tweezer based on a single focused ultrasound beam
provides a simple method for simultaneous manipulation of microparticles (Rayleigh particles).
By performing computational simulations based on the paraxial approximation method, 
it has been demonstrated that a focused piezoelectric transducer  operating at $\unit[1]{MHz}$ with an intensity of $\unit[3.3]{kW/m^2}$
can axially trap microdroplets in the transducer pre-focal zone. 
The trapping radiation force is  in the nanonewtons range.
We can further notice in Fig.~\ref{fig:axial_force} that off-axis stable points are also present in the pre-focal region.

Unlike entrapment in the focal region, the stability of traps in the pre-focal zone is not affected by  
the absorption and the scattering radiation forces.
In principle, any less dense and more compressible microdroplet with respect to the host liquid can be  trapped
in the pre-focal zone of the transducer.
Moreover, numerical simulations of the particle dynamics in a viscous fluid have shown that the microdroplets 
are indeed trapped in the pre-focal zone.
We emphasize that these results depends on the paraxial approximation used to derive the acoustic fields
generated by the transducer.

A numerical simulation of the wave pressure generated by  the focused transducer 
was performed based on the finite element method (FEM).
After calculating the gradient  force on an oil microdroplet, we also found axial trapping points
in the pre-focal region. 
The methods have good agreement in the region $z>\unit[20]{mm}$.

If experimental investigations confirm the theoretical predictions exposed here,
the proposed method might be  an advance in microparticle handling devices
based on piezoelectric transducers.

\appendix
\section{Axial momentum flux  and time-averaged intensity}
Let us analyze the divergence of the momentum flux  $\textrm{Im}[\nabla \cdot \rho_0{\bm v}_\textrm{in} {\bm v }_\textrm{in}^*]$  in Eqs.~(\ref{fscat1}) and (\ref{fabs}).
Since the paraxial beam described in Eq.~(\ref{qrz2}) is axisymmetric,
 the azimuthal component of the fluid element velocity is zero, $v_{\textrm{in},\varphi} = 0$.
Hence,  the divergence of the momentum flux term (second-rank tensor) 
of an axisymmetric beam, in cylindrical coordinates, is given by~\cite{lebedev2010}
\begin{align}
\nonumber
\nabla \cdot \rho_0{\bm v}_\textrm{in} {\bm v }_\textrm{in}^* &=
\rho_0 \biggl[ \left[
\partial_\varrho\left( v_{\textrm{in},\varrho}v_{\textrm{in},\varrho}^*\right)
+ \partial_z \left( v_{\textrm{in},\varrho}v_{\textrm{in},z}^*\right)+ \frac{v_{\textrm{in},\varrho}v_{\textrm{in},\varrho}^*}{\varrho}\right]\bm{e}_\varrho\\ 
&+ \left[ \partial_\varrho \left( v_{\textrm{in},\varrho}v_{\textrm{in},z}^*\right) +
\frac{v_{\textrm{in},\varrho}v_{\textrm{in},z}^*}{\varrho}  +
\partial_z \left( v_{\textrm{in},z}v_{\textrm{in},z}^*\right) 
\right] \bm{e}_z \biggr].
\end{align}
By taking the imaginary-part of this expression, we get
\begin{equation}
\textrm{Im}[\nabla \cdot \rho_0{\bm v}_\textrm{in} {\bm v }_\textrm{in}^*] =
\rho_0 \biggl[ 
\partial_z \left( v_{\textrm{in},\varrho}v_{\textrm{in},z}^*\right)\bm{e}_\varrho+
\left [ \partial_\varrho \left( v_{\textrm{in},\varrho}v_{\textrm{in},z}^*\right) +
\frac{v_{\textrm{in},\varrho}v_{\textrm{in},z}^*}{r} \right] \bm{e}_z \biggr].
\label{Im_r_vv2}
\end{equation}
Referring to Eq.~(\ref{vp2}) and considering $\varrho=0$, 
we find that the transverse component of the fluid element velocity is $v_{\textrm{in},\varrho}(0,z)=0$.
Thus, in the axial direction Eq.~(\ref{Im_r_vv2}) becomes
\begin{equation}
\textrm{Im}[\nabla \cdot \rho_0{\bm v}_\textrm{in} {\bm v }_\textrm{in}^*(0,z)] =\textrm{Im}[\partial_\varrho \left( v_{\textrm{in},\varrho}v_{\textrm{in},z}^*\right)_{\varrho=0}] \bm{e}_z.
\end{equation}
With aid of \textsc{Mathematica} software~\cite{mathematica10} and using  Eq.~(\ref{vp2}), we obtain 
\begin{align}
\nonumber
\textrm{Im}[\nabla \cdot \rho_0{\bm v}_\textrm{in} {\bm v }_\textrm{in}^*(0,z)] &=
\frac{\rho_0 v_0^2 z_0^2}{2z^2(z-z_0)^3}
\biggl[ 8 z^2 \sin^2\left[\frac{k b^2(z-z_0)}{4 z z_0}\right] \\
&- k b^2(z-z_0) \sin\left[\frac{k b^2(z-z_0)}{2 z z_0}\right]
\biggr]\bm{e}_z.
\label{div_flux}
\end{align}

Another important quantity in the radiation force analysis is the axial time-averaged intensity, which is given
for the focused ultrasound beam by
\begin{align}
\nonumber
\overline{\bm{I}}(0,z) &= 
\frac{1}{2}
\textrm{Re}[p_\textrm{in}(0,z) v_{\textrm{in},z}^*(0,z)]\bm{e}_z \\
&= \frac{\rho_0 v_0^2 c_0}{2} \frac{ (4z^2-b^2)z_0^2}{z^2(z-z_0)^2}
\sin^2\left[\frac{k b^2(z-z_0)}{4 z z_0}\right]\bm{e}_z.
\label{Iz2}
\end{align}

\section*{Acknowledgements}
This work was partially supported by grants 481284/2012-5 and 303783/2013-3 CNPq (Brazilian agency).

%\pagebreak
%\bibliographystyle{elsart-num}
%\bibliography{tweezer}

\end{document}